\documentclass[%
 reprint,
 amsmath,amssymb,
 aps,
prb,
superscriptaddress,
]{revtex4-1}

\usepackage{graphicx}
\usepackage{dcolumn}
\usepackage{bm}
\usepackage{hhline}
\usepackage[utf8]{inputenc}
\usepackage[english]{babel}
\begin{document}

\preprint{APS/123-QED}

\title{\textit{Ab-initio} design of new Heusler materials for thermoelectric applications}
\author{Mohd Zeeshan}
\affiliation{Indian Institute of Technology Roorkee, Department of Chemistry, Roorkee 247667, Uttarakhand, India}
\author{Harish K. Singh}
\affiliation{TU Darmstadt, Theory of Magnetic Materials, Department of Materials- and Earth Science, Alarich-Weiss-Str.~16, 64287 Darmstadt, Germany}
\author{Jeroen van den Brink}
\affiliation{Institute for Theoretical Solid State Physics, IFW Dresden, Helmholtzstrasse 20, 01069 Dresden, Germany}
\author{Hem C. Kandpal}
\affiliation{Indian Institute of Technology Roorkee, Department of Chemistry, Roorkee 247667, Uttarakhand, India}
\affiliation{Institute for Theoretical Solid State Physics, IFW Dresden, Helmholtzstrasse 20, 01069 Dresden, Germany}

\date{\today}

\begin{abstract}
In search of new prospects for thermoelectric materials, using \textit{ab-initio} calculations and semi-classical Boltzmann theory, we have systematically investigated the electronic structure and transport properties of 18-valence electron count cobalt based half-Heusler alloys with prime focus on CoVSn, CoNbSn, CoTaSn, CoMoIn, and CoWIn. The effect of doping on transport properties has been studied under the rigid band approximation. The maximum power factor, S$^2\sigma$, for all systems is obtained on hole doping and is comparable to the existing thermoelectric material CoTiSb. The stability of all the systems is verified by phonon calculations. Based on our calculations, we suggest that CoVSn, CoNbSn, CoTaSn, CoMoIn and CoWIn could be potential candidates for high temperature thermoelectric materials.

\end{abstract}

\maketitle


\section{Introduction}
Thermoelectric (TE) materials have seen a huge upsurge in recent years owing to their potential applications in power generation from waste heat\cite{Bell2008,Fu2015}. However, the low efficiency of TE materials has been the prime obstruction in replacing the traditional methods of power generation\cite{Biswas2012,Zhao2014}. The efficiency of the TE material is given by a dimensionless quantity called the figure of merit \textit{ZT}, given by \textit{ZT = S$^2\sigma$T/$\kappa$}, where \textit{S} is Seebeck coefficient, \textit{$\sigma$} is electrical conductivity, \textit{T} is absolute temperature, and \textit{$\kappa$} is thermal conductivity. The thermal conductivity comprises of two parts, $\kappa$ = $\kappa_e$ + $\kappa_l$, where $\kappa_e$ represents thermal conductivity from electrons and \textit{$\kappa_l$} from the lattice. With recent advances in synthesis methods, characterization techniques and powerful computational tools, the \textit{ZT} value has seen a progressive increment but is yet to meet the threshold value for commercial applications\cite{Poon2011,Joshi2014}.

The foremost criterion for the choice of TE materials is a narrow band gap semiconductor\cite{Graf2011,Snyder2008,Bos2014,Tritt2011}. Also, the viable application of thermoelectric materials requires high temperature sustainable, low cost, and non-toxic materials\cite{Wang2013,Tan2017,Cerretti2017,Ge2016}. One important class of TE materials, which meets the above criteria, is half-Heusler (hH) alloys. The growing interest in hH alloys is partly driven by their robust properties, \textit{viz.} thermal stability, mechanical strength, low cost, non-toxicity, and semiconducting behavior\cite{Misra2014,Poon2001,Chen2013,Huang2016,Xie2012}.

One remarkable property which makes hH alloys particularly interesting, as mentioned before, is the narrow band gap semiconducting behavior of 18-valence electron count (VEC) ternary hH alloys\cite{Young2000,Culp2008,Casper2009,Casper2012}. Generally, the 18-VEC hH alloys are found to be in agreement with the Slater-Pauling Rule, \textit{i.e.} \textit{M} = \textit{N$_v$} – 18, where \textit{N$_v$} = valence electrons and \textit{M} is the magnetic moment\cite{Galanakis2002}. Most evidently, cobalt based 18-VEC hH alloys follow the Slater Pauling Rule, \textit{i.e.} non-magnetic semiconductors\cite{Fecher2006}. For instance, CoTiSb has \textit{N$_v$} = 18, \textit{M} = 0, and is a non-magnetic semiconductor\cite{Galanakis2006,Kubler1984,Felser2007,Galanakis2002}. In the past two decades, a large number of hH alloys have been reported to exhibit substantial TE properties comparable to conventional Bi$_2$Te$_3$ and PbTe based materials\cite{Fang2017,Zhang2016,Yu2009,Joshi2011,Yan2012,Sakurada2005,Li2016,Fang2016}. S. J. Poon \textit{et al.} reported that the \textit{n}-type Hf$_{0.6}$Zr$_{0.4}$NiSn$_{0.995}$Sb$_{0.005}$ hH alloy and \textit{p}-type Hf$_{0.3}$Zr$_{0.7}$CoSn$_{0.3}$Sb$_{0.7}$/nano-ZrO$_2$ composites achieved \textit{ZT} = 1.05 and 0.8 near 900 – 1000~K, respectively\cite{Poon2011}. In another work, Joshi \textit{et al.} reported a high \textit{ZT} value of $\sim$1 at 700$^{\circ}$~C for a nanostructured \textit{p}-type Nb$_{0.6}$Ti$_{0.4}$FeSb$_{0.95}$Sn$_{0.05}$ composition\cite{Joshi2014}.  

More recently, Zunger and coworkers, with the aid of first-principles calculations, systematically investigated the cobalt, rhodium, and iridium based 18-VEC hH alloys and revealed some new thermodynamically stable systems\cite{Zakutayev2013}. In addition, for the first time, they experimentally realized the CoTaSn system. This is the motivation to search for the missing 18-VEC hH alloys. Till now, a large number of hH alloys have been documented in Inorganic Crystal Structure Database (ICSD) and still there exists a wide horizon to search for new potential hH candidates\cite{ICSD}. In the quest of the same, we start screening the cobalt based 18-VEC hH alloys. To the best of our knowledge, CoTiSb, CoZrSb, CoHfSb, CoVSn, CoNbSn, and CoTaSn are well explored in theory and experiment. However, the credibility of CoVSn and CoTaSn for a potential TE material is yet to be tested. From the reported literature so far, we discovered that the CoCrIn, CoMoIn, and CoWIn are still missing from the timeline of TE materials.  

\begin{figure}
\centering
 \includegraphics[scale=0.30]{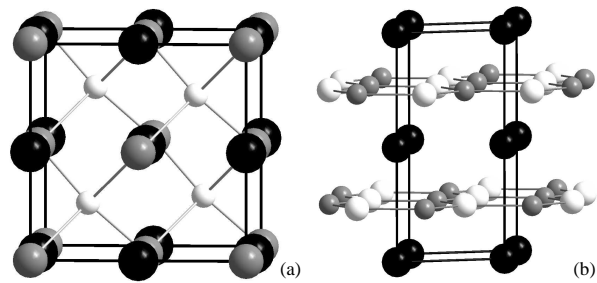}
 \caption{Crystal structure of \textit{XYZ} hH alloy (a) in \textit{F$\bar{4}$3m} symmetry and (b) in \textit{P6$_3$/mmc} symmetry. White, black, and grey spheres represent \textit{X}, \textit{Y}, and \textit{Z}, respectively. \textit{X-Z} crystallizes in zinc-blende type sub-lattice in (a) and planar honeycomb sublattice in (b).}
\label{crystal}
\end{figure}

The subject of the present work is to systematically investigate the thermodynamic stability, dynamical stability, and explore the electronic transport properties of new plausible cobalt based hH alloys in both cubic (\textit{F$\bar{4}$3m}) and hexagonal (\textit{P6$_3$/mmc}) space groups with the help of first-principles calculations. The idea behind the hexagonal crystal structure stems from the earlier work of Y. Noda and F. Casper. F. Casper \textit{et al.} searched for the hexagonal analogues of half-metallic hH \textit{XYZ} alloys\cite{Casper2008}. In another work, Y. Noda \textit{et al.} studied a phase transition from cubic-CoVSb to hexagonal-CoVSb on applying a high pressure of 5~GPa\cite{Noda1979}. The cubic-hexagonal phase transition is somewhat analogous to a diamond-graphite phase transition. Here, the diamond like tetrahedral environment of \textit{XZ} collapses into graphite like honey-comb sub-lattice (Fig.~\ref{crystal}). The pressure induced cubic-hexagonal phase transition may have pronounced effect on the nature of chemical bonding, structural, and electronic properties of materials. For instance, on the application of pressure, insulators may become metallic or vice versa\cite{Manjon2009}. The complete redistribution of structural and electronic properties may be useful for improving the efficiency of existing TE materials.
 
The present work is categorized into three parts: i) Computational Details: This section includes a brief description of the computational tools utilized for carrying out present work. ii) Results: Within this section, we will explain the crystal structure of hH alloys, structural optimization of Co\textit{YZ} systems (\textit{Y} = Ti, Zr, Hf, V, Nb, Ta, Cr, Mo, W \& \textit{Z} = Sb, Sn, In) in \textit{F$\bar{4}$3m} and \textit{P6$_3$/mmc} space group, nature of band gap, crystal structure stability (phonon calculations), electronic structure (Band and DOS) calculations, and finally transport properties and their behavior with doping concentration at different temperatures. iii) Discussion and Conclusions: Here, we will discuss the importance of doping in hH alloys and emphasize on the interplay of theory and experiment for designing new potential TE materials.

\section{Computational Details}
In the present work, we use a combination of two different first-principles density functional theory (DFT) codes: the full-potential linear augmented plane wave method (FLAPW)\cite{Singh2006} implemented in WIEN2k\cite{Blaha2001} and the plane-wave pseudopotential approach implemented in Quantum ESPRESSO package\cite{Giannozzi2009}. The former is used to obtain equilibrium lattice constants, electronic structure, and transport properties, and the latter method to confirm the structure stability by determining the phonon spectrum.

The FLAPW calculations are performed using a modified Perdew-Burke-Ernzerhof (PBEsol correlation)\cite{Perdew2008} implementation of the generalized gradient approximation (GGA). For all the calculations, the scalar relativistic approximation is used. The muffin-tin radii (RMTs) are taken in the range 2.3-2.6 Bohr radii for all the atoms. RMT x kmax = 7 is used as the plane wave cutoff. The self-consistent calculations were employed using 64000 \textit{k}-points in the full Brillouin zone. The energy and charge convergence criterion was set to 10$^{-6}$ and 10$^{-5}$, respectively. Calculations are done for different volumes in the cubic structure (\textit{F$\bar{4}$3m}), as well as in the hypothetical hexagonal structure (\textit{P6$_3$/mmc}).

The electronic transport properties are calculated using the Boltzmann theory\cite{Allen} and relaxation time approximation as implemented in the BoltzTraP code\cite{Madsen2006}. The electrical conductivity and power factor (PF) are calculated with respect to time relaxation, \textit{$\tau$}, the Seebeck coefficient is independent of \textit{$\tau$}. The relaxation time was calculated by fitting the available experimental data with theoretical data. We have used this approach in evaluating the electronic transport properties of our systems.  

In the plane-wave pseudopotential approach, we use scalar-relativistic, norm-conserving pseudopotentials for a plane-wave cutoff energy of 100~Ry. The exchange-correlation energy functional was evaluated within the generalized gradient approximation (GGA), using the Perdew-Burke-Ernzerhof parametrization\cite{Perdew1996}, and the Brillouin zone is sampled with a 20$\times$20$\times$20 mesh of Monkhorst-Pack \textit{k}-points. The calculations are performed on a 2$\times$2$\times$2 \textit{q}-mesh in the phonon Brillouin zone.

\section{Results}
In order to exploit the overlying transport properties of any material, one must have the better understanding of the underlying crystal structure and electronic structure. Therefore, in the coming sub-section, we focus on the detailed analysis of crystal structure and structural optimization of 9 Co-based hH alloys (CoTiSb, CoZrSb, CoHfSb, CoVSn, CoNbSn, CoTaSn, CoCrIn, CoMoIn, and CoWIn) under investigation. In the following sub-sections, we will discuss the electronic structure of above-mentioned systems and transport properties of the selected systems.

\subsection{Crystal Structure}
The family of hH alloys, \textit{XYZ} (\textit{X} and \textit{Y} = transition metals and \textit{Z} = main group element) crystallizes in non-centrosymmetric MgAgAs type structure (Space Group \textit{F$\bar{4}$3m}) (Fig.~\ref{crystal}). The crystal structure can be visualized as the \textit{XZ} zinc-blende structure in a diamond like network, stuffed with \textit{Y} atoms. The \textit{X} atoms form two identical tetrahedra, both with \textit{Y} and \textit{Z}, respectively. However, \textit{Y} and \textit{Z} both form an octahedra with each other and a tetrahedra with \textit{X}, respectively.  Another view point of the crystal structure can be regarded as the stuffed combination of a rock-salt type (NaCl) structure and zinc-blende (ZnS) type structure. The most electropositive transition element \textit{Y} (Ti, Zr, Hf, V, Nb, Ta, Cr, Mo, In) and the most electronegative main group element \textit{Z} (Sb, Sn, In) crystallizes in rock-salt type sub-lattice, whereas the intermediate electronegative transition element \textit{X} (Co) and most electronegative transition element \textit{Z} crystallizes in zinc-blende type sub-lattice.  The strong covalent character between \textit{X} and \textit{Z} is considered to be the prima facie for semiconducting band gap in 18-VEC hH alloys\cite{Casper2008}. The Wyckoff positions for \textit{X}, \textit{Y}, and \textit{Z} are (1/4, 1/4, 1/4), (0, 0, 0), and (1/2, 1/2, 1/2), respectively. The remaining (3/4, 3/4, 3/4) sites are vacant which makes hH alloys favorable for doping\cite{Graf2011,Casper2012,Casper2009,Galanakis2002,Casper2008,Kandpal2006,Nowotny1950,Offernes2007,Jung2000,Tobola2000,Pierre1997}.

\subsection{Structural Optimization}
From here onwards, we classify CoTiSb, CoZrSb, CoHfSb as Ti-group; CoVSn, CoNbSn, CoTaSn as V-group; and CoCrIn, CoMoIn, CoWIn as Cr-group. In order to establish the ground state properties, we optimized all 9 systems in both cubic and hexagonal framework. We minimized the total energy as a function of volume, fitted with Birch-Murnaghan equation\cite{Birch1947}, for cubic systems in \textit{F$\bar{4}$3m} symmetry. And for hexagonal systems, we minimized the energy as a function of volume and \textit{c/a} parameter in \textit{P6$_3$/mmc} symmetry. The calculated lattice parameters for both cubic and hexagonal systems are listed in Table~\ref{tab1}. The calculated lattice parameters for cubic Ti-group and V-group are in good agreement with experimental values in parenthesis. The reliability of our calculations lies in the fact that the discrepancy between calculated and experimental lattice parameters lies in the range of 0.69-1.01\%\/, except for CoVSn system (4.3\%\/). However, the calculated lattice parameter for CoVSn by M. Hichour \textit{et al.}\cite{Hichour2012} and M. Ameri \textit{et al.}\cite{Ameri2013} also showed the same discrepancy. The trend to be followed for Ti-group and V-group, from Table~\ref{tab1}, is the increase in lattice parameter on going from \textit{3d} to \textit{4d} element (Ti to Zr and V to Nb) and subsequent decrease on going from \textit{4d} to \textit{5d} element (Zr to Hf and Nb to Ta). The corresponding increase in lattice parameter in going from first to the second member of the same family is quite apparent. But the increase in lattice parameter, in going from CoZrSb to CoHfsb and CoVSn to CoNbSn, can be attributed to lanthanide contraction\cite{Housecroft2004,Cotton1988}. However, a regular increase in lattice parameter was observed for the Cr-group. We find that GGA interestingly underestimates the lattice parameter on average by 1\%\/. 

\begin{table}
\begin{tabular}{l|c|c|c|c|c}
\hline
Compound	&	\multicolumn{3}{c|}{$F\bar{4}3m$}		&	\multicolumn{2}{c}{$P6_3/mmc$} \\
\hline
		&	$a$ (\AA)	&	$E_g$ (eV)	&	$B_0$ (GPa)	& $a$ (\AA)	& $c$ (\AA) \\	
\hline
CoTiSb	&	5.82 (5.8818)	&	1.11 (0.19)	&	156.4	&	4.344  &	5.509 \\        
CoZrSb	&	6.03 (6.0676)	&	1.08 (0.14)	&	148.6	&	4.394  &	6.072 \\
CoHfSb	&	5.99 (6.0383)	&	1.14 (0.07)	&	154.8	&	4.394  &	5.958 \\
\hline
CoVSn	&	5.73 (5.9800)	&	0.66 (0.75)	&	166.5	&	4.330	&	5.205 \\
CoNbSn	&	5.90 (5.9559)	&	1.02 (1.00)	&	170.6	&	4.427	&	5.518 \\
CoTaSn	&	5.89 (5.9400)	&	1.06 (1.30)	&	180.4	&	4.415	&	5.536 \\
\hline
CoCrIn	&	5.68 (-)	&	0.00 (-)	&	161.5	&	4.339	&	4.946 \\
CoMoIn	&	5.80 (-)	&	0.04 (-)	&	178.4	&	4.444	&	5.150 \\
CoWIn	&	5.80 (-)	&	0.51 (-)	&	193.7	&	4.429	&	5.214 \\
\hline
\end{tabular}             
\caption{Optimized cell parameters, band gap, and bulk modulus for Co\textit{YZ} hH alloys in cubic symmetry and optimized cell parameters for Co\textit{YZ} hH alloys in hexagonal symmetry. The corresponding experimental values are given in parentheses.}
\label{tab1}
\end{table}	

From the view-point of stability, all but CoVSn and CoTaSn cubic systems are found to be more stable than their hexagonal counterparts. To our surprise, present calculations reveal that hexagonal-CoVSn is more stable than cubic-CoVSn by 0.68~eV and hexagonal-CoNbSn is more stable than cubic-CoNbSn by 0.59~eV per formula unit. However, there is no experimental evidence for the same, yet. Here, we predict that CoVSn and CoNbSn could be synthesized in hexagonal symmetry under certain conditions.

\begin{figure}
\centering
 \includegraphics[scale=0.40]{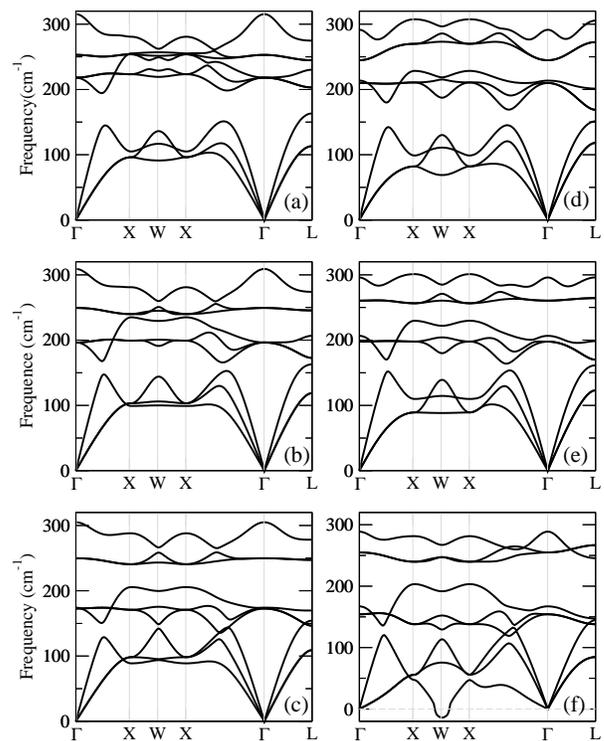}
 \caption{(a-f) shows phonon dispersion curves for CoVSn, CoNbSn, CoTaSn, CoCrIn, CoMoIn, and CoWIn, respectively in \textit{F$\bar4$3m} symmetry.}
\label{phonon}
\end{figure}

The prototype structure of hexagonal counterpart of \textit{XYZ} is ZrBeSi (\textit{P63/mmc}) structure (Fig.~\ref{crystal}). As discussed before, few cubic 18-VEC hH alloys like CoVSb and FeVSb are found to exist in this space group under certain conditions. Here, \textit{X} and \textit{Z} form planar graphite like honey-comb sub-lattice. The Wyckoff positions for \textit{X}, \textit{Y} and \textit{Z} are (1/3, 2/3, 3/4), (0, 0, 0), and (1/3, 2/3, 1/4), respectively\cite{Casper2008,Bennett2012}.

The calculated band gaps of all cubic systems, except CoCrIn, shows semiconducting behavior and are listed in Table~\ref{tab1}. The band gap values for cubic systems lies in the range of 0.03 - 1.13~eV which is good for TE materials. The calculated band gap values for Ti-group and V-group are in good agreement with previously calculated values. However, the experimental band gap values for CoTiSb (0.19~eV) and CoHfSb (0.14~eV) are very small in comparison to calculated values. The possible explanation, as suggested by T. Sekimoto \textit{et al.}, could be the deviation from stoichiometric composition and anti-site disordering \cite{Sekimoto2005}. There was no previous evidence of calculated band gap for CoZrSb and experimental band gap for V-group. In Cr-group, CoMoIn and CoWIn show narrow band gap whereas CoCrIn is metallic in nature. All systems in hexagonal symmetry also show metallic behavior, \textit{i.e.} insulator to metallic transition on applying hydrostatic pressure to the cubic system. Since the metallic systems are not good choices for TE materials, we discard the cubic-CoCrIn and all hexagonal systems for further evaluation. For further study, we select CoVSn, CoNbSn, CoTaSn, CoMoIn, and CoWIn systems.

The bulk modulus, listed in Table~\ref{tab1}, gives the idea of the stiffness of the material. The calculated bulk modulus of all the systems is comparable to that of steel and is an indication of a strong material. From Table~\ref{tab1}, it is easy to notice the strength of CoTaSn, CoMoIn, and CoWIn. The stiffness of these systems is a good indication since the hH alloys are desired to work under robust conditions.

\subsection{Phonon Calculations}
The most stable structures obtained from the optimization were tested by examination of their dynamic stability with phonon calculations. We performed a two-step phonon calculation. First, we optimized the crystal structure of V-group and Cr-group by using Quantum ESPRESSO, based on DFT and plane-wave pseudo-potential method. The optimized results were in good agreement with our WIEN2k calculations. Next, we calculated the phonon dispersion by using the density function perturbation theory (DFPT) implemented in Quantum ESPRESSO. The calculations were performed on a 2$\times$2$\times$2 mesh in the phonon Brillouin zone, and force constants in real space derived from this input are used to interpolate between \textit{q}-points and to obtain the continuous branches of the phonon band structure.

Phonons can be considered as normal modes or quantum of vibrations in crystal and serve the purpose of crystal structure stability. Phonon stability lies in the fact that the frequency for each phonon should be a real quantity and not imaginary \cite{Eliott2006,Togo2015}. As can be seen from Fig.~\ref{phonon}, there are no imaginary frequencies throughout the Brillouin Zone for V-group and Cr-group, except CoWIn, where imaginary frequencies are observed along W-direction. However, the imaginary frequencies up to 10~cm$^{-1}$ are not of much concern and can be removed by employing anharmonic approximations\cite{Hellman2011}. Thus, we ensure the dynamic stability of V-group and Cr-group alloys and proceed next to see how the electronic structure behaves in these systems.

\subsection{Electronic Structure}
\begin{figure}
\centering
 \includegraphics[scale=0.40]{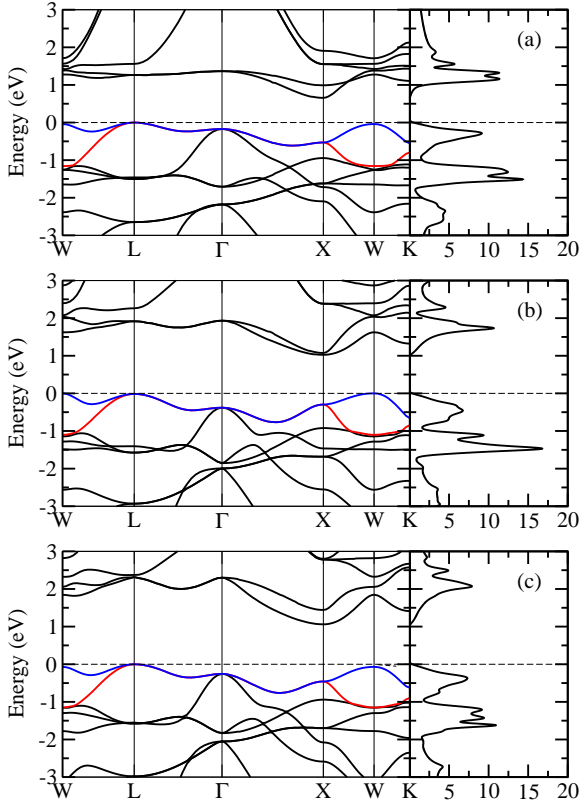}
 \caption{(a-c) shows calculated electronic structures for CoVSn, CoNbSn, and CoTaSn, respectively in \textit{F$\bar{4}$3m} symmetry. The top of the valence band is taken as zero on the energy axis.}
\label{dos1}
\end{figure}

\begin{figure}
\centering
 \includegraphics[scale=0.40]{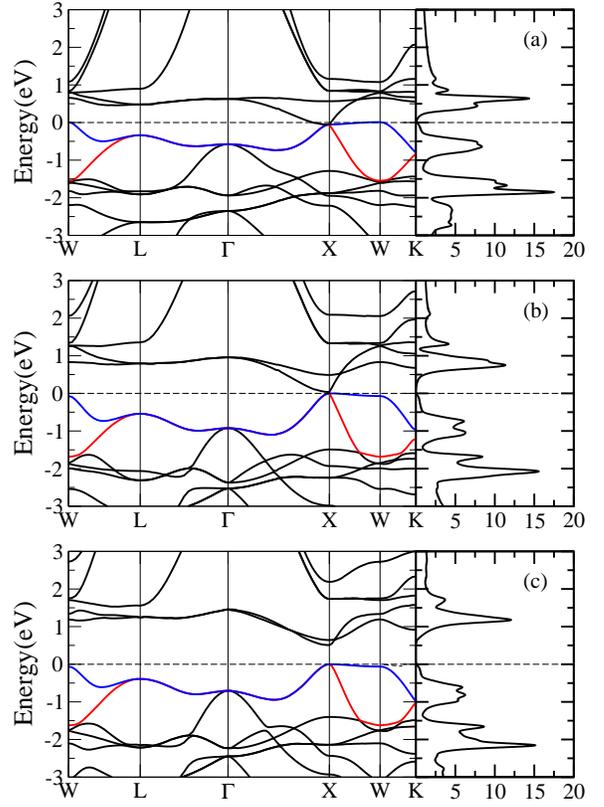}
 \caption{(a-c) shows calculated electronic structures for CoCrIn, CoMoIn, and CoWIn respectively in \textit{F$\bar{4}$3m} symmetry. The top of the valence band is taken as zero on the energy axis.}
\label{dos2}
\end{figure}

In the present section, we discuss the electronic structure of V-group and Cr-group. The band structure and DOSs are shown in Fig.~\ref{dos1} and Fig.~\ref{dos2}. Taking transport properties into account, some gross features from band structure and DOSs plot are discussed here. 

i) The states (not shown for clarity) participating at valence band maxima (VBM) and conduction band minima (CBM) for all the systems, mostly originates from the \textit{d-d} mixing between \textit{X} and \textit{Y} atoms, along with some low lying \textit{p}-states of \textit{Z} atom. The \textit{d-d} mixing may vary depending on the combination of \textit{X-Y}. ii) The interplay of degeneracy and dispersion of bands at VBM plays an important role in transport properties. The VBM of V-group, except for CoNbSn, lies at L-point and is 2-fold degenerate. The VBM of CoNbSn lies at L- and W-points. Here, both L- and W-points can contribute towards charge carrier transition, thereby enhancing the TE properties. The flat bands in L-$\Gamma$ region are reflective of \textit{d-d} mixing between \textit{X-Y} atoms. 2-fold degeneracy and flat band correspond to the heavier mass of charge carriers which enhances Seebeck coefficient. However, heavy charge carriers at VBM lead to a reduction in electrical conductivity. The contribution towards electrical conductivity may come from low lying bands below VBM, having low effective mass of charge carriers. The VBM of Cr-group lies at X-point and is also 2-fold degenerate. The flatter bands in this group appear in X-W region and are less dispersed as compared to V-group, indicative of low thermopower, \textit{S}\cite{Yadav2015,Lee2011,Yang2008,Parker2010}. iii) The nature of band gap of V-group is indirect (L-X) whereas that of Cr-group is direct at X-point. In the case of indirect band-gap, the VBM and CBM occur at different wave vector (\textit{k}) and hence, a change in momentum is required for charge carrier transition. This change in momentum is provided by the crystal lattice in the form of phonons, thereby, increasing the thermal conductivity\cite{Ashcroft2014}. Since in most hH alloys, the lattice thermal conductivity part dominates electronic thermal conductivity; a significant reduction in \textit{ZT} value is expected\cite{Chen2013}. Therefore, the direct band gap semiconductors provided small total thermal conductivity, are better candidates for achieving higher \textit{ZT}. The bulk modulus, dynamic thermal stability, and direct band gap nature, collectively support the TE potential of CoMoIn and CoWIn. Thus, it would be interesting to see how their transport properties will respond.

\subsection{Transport Properties}
\begin{figure}
\centering
 \includegraphics[scale=0.40]{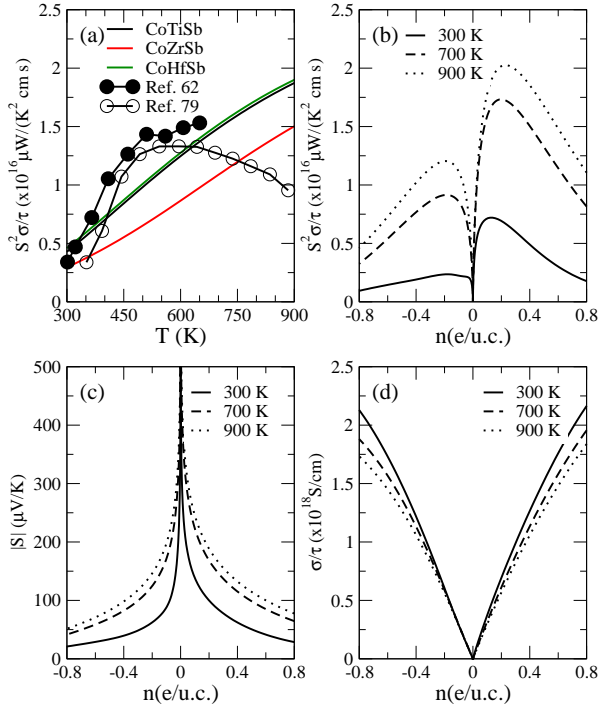}
 \caption{(a) shows the comparison of calculated and experimental power factor as a function of temperature for CoTiSb, along with calculated  power factor for CoZrSb and CoHfSb. (b-d) shows the trend of power factor, Seebeck coefficient, and electrical conductivity on doping CoTiSb at 300~K, 700~K, and 900~K, respectively. Power factor and electrical conductivity are plotted with respect to relaxation time.}
\label{transport1}
\end{figure}

In this section, using rigid band approximation (RBA), semi-classical Boltzmann theory, and constant relaxation time (\textit{$\tau$}), we calculate transport properties and predict the optimal doping concentrations for attaining maximum PF. This would encourage the experimentalists to choose a narrow range for varying doping levels. The RBA assumes that on doping a system, the Fermi level moves up or down without any alteration in band structure. Therefore, a single band structure calculation is sufficient for all doping concentrations. The validity of RBA holds good for low doping levels and has been widely used by many groups\cite{Madsen2006,Lee2011,MadsenJACS2006,Chaput2005,Jodin2004}. Fig.~\ref{transport1} and Fig.~\ref{doping} shows the calculated transport properties for CoTiSb (in comparison with experiment) and V-group and Cr-group, respectively. Before discussing individual systems, first, we discuss the common features of transport properties of CoTiSb, V-group, and Cr-group. Seebeck coefficient of all the systems is maximum when the Fermi level is close to the middle of the band gap and drops almost exponentially with doping. However, exactly opposite trend is observed for electrical conductivity. Electrical conductivity is very low when Fermi level is close to the band gap but increases rapidly on doping. When Fermi level shifts towards the VBM or CBM, the density of states increases at the Fermi level, thereby increasing the electrical conductivity and lowering the Seebeck coefficient\cite{Lee2011,Yang2008,Parker2010}.

As per Mott equation\cite{Heremans2008,Cutler1969}, for heavily doped systems, Seebeck coefficient decreases with increasing carrier concentration. Therefore, a TE material should be tuned in such a way that it maintains a high Seebeck coefficient without any significant reduction in electrical conductivity. The combined effect of the dependence of Seebeck and electrical conductivity on doping reflects that the maximum PF is obtained when the Fermi level is near the band edge. Next, we proceed to discuss the individual systems. 

First, to check the reliability of our calculations, we compare our results with the well-known CoTiSb system. We choose CoTiSb as the reference since our interest is in unreported Co\textit{YZ} systems. Also, out of 9 Co\textit{YZ} systems, CoTiSb is most studied in theory and experiment by different groups\cite{Galanakis2002,Sekimoto2005,Zhou2007,Birkel2012,Ouardi2012,Wu2007,Zhou2005,Sekimoto2006,Wu2009}. The calculated transport properties for CoTiSb are shown in Fig.~\ref{transport1}(a-d). Fig.~\ref{transport1}(a) shows the behavior of calculated and experimental PF/\textit{$\tau$} with temperature. In order to compare the theory with experiment, one has to multiply the calculated PF by relaxation time, \textit{$\tau$}. Here, we employed the reverse approach and divided the reported PF by \textit{$\tau$} to obtain the experimental PF/\textit{$\tau$}. Since there have been no prior calculated \textit{$\tau$} values for CoTiSb, we resorted to reported electrical conductivity values. In a crude approximation, we have taken \textit{$\tau$} = \textit{$\sigma$}$_{exp}$/\textit{$\sigma$}$_{cal}$ and found that \textit{$\tau$} is of the order of 10$^{-16}$ s. Incorporating an average value of \textit{$\tau$} = 2 x 10$^{-16}$ into PF, reported by Sekimoto \textit{et al.} \cite{Sekimoto2005} and Birkel \textit{et al.}\cite{Birkel2012}, we obtain a nice agreement between calculated and reported PF/\textit{$\tau$} (Fig.~\ref{transport1}(a)). This encourages us to study the CoTiSb with different doping concentrations, and compare with previously calculated and experimental values. 

Fig.~\ref{transport1}(b) shows PF/\textit{$\tau$} for different doping concentrations at 300~K, 700~K and 900~K, respectively. We have chosen high temperatures because the practical applicability of hH alloys requires high-temperature sustainability. For all temperatures, PF increases substantially on both electron and hole doping and then drops at higher doping levels. However, hole doping dominates for all temperatures. The improvement in PF for both \textit{n}- and \textit{p}-type doping in CoTiSb has already been reported by many groups\cite{Zhou2007,Ouardi2012,Wu2007,Zhou2005,Sekimoto2006,Wu2009}. The peak value of calculated PF/\textit{$\tau$} at 300~K is obtained for 0.12 hole doping, at 700~K for 0.20 hole doping, and at 900~K for 0.22 hole doping per unit cell. These doping concentrations are quite pragmatic and could be realized experimentally. For instance, the 0.12 hole doping for 300~K could be realized by substituting 12\% Sb by Sn or 12\% Co by Fe. This is in stark agreement with previously calculated and reported values. J. Yang \textit{et al.} calculated that 15\%\/ hole doping could lead to maximum PF at 300~K\cite{Yang2008}. T. Wu \textit{et al.} reported a maximum PF of 23 $\mu$W/K$^2$~cm at 850~K for Co$_{0.85}$Fe$_{0.15}$TiSb\cite{Wu2007}. In our calculations, the maximum calculated PF/\textit{$\tau$} is 2~$\times$~10$^{16}$~$\mu$W/K$^2$~cm~s at 900~K for 0.22 hole doping per unit cell. The corresponding Seebeck coefficient, \textit{S}, and electrical conductivity with respect to relaxation time, $\sigma$/$\tau$, are 173 $\mu$V/K and 0.63 $\times$ 10$^{18}$ S/cm~s, respectively. The value of Seebeck coefficient lies well in the range of reported Seebeck coefficient for hole doped CoTiSb\cite{Wu2007,Sekimoto2006,Wu2009}.

\begin{figure}
\centering
 \includegraphics[scale=0.35]{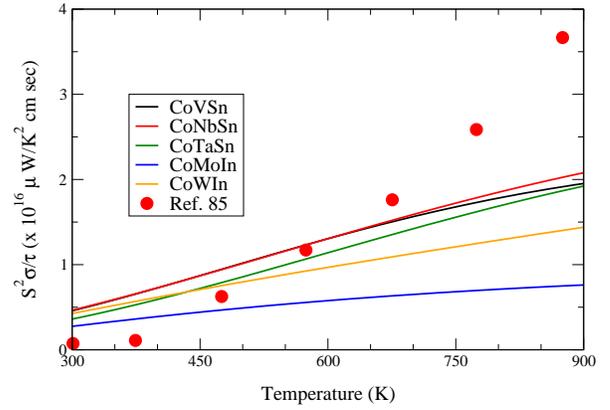}
 \caption{shows the calculated power factor, with respect to relaxation time, as a function of temperature for V-group and Cr-group, along with experimental power factor for CoNbSn.}
\label{transport2}
\end{figure}

Once we established that our calculated transport properties for CoTiSb are in accordance with reported work, we proceed next to V-group and Cr-group. As per our findings, CoVSn, CoNbSn, and CoTaSn are reported in theory and experiment. However, the transport properties of CoVSn and CoTaSn are yet to be studied in detail. Before exploring the transport properties of CoVSn, CoTaSn, and Cr-group, we begin with analyzing our calculated transported properties of CoNbSn with reported values.

Fig.~\ref{transport2} shows the PF/\textit{$\tau$} as a function of temperature for V-group, Cr-group, and reported CoNbSn. Once again, to correlate the calculated and reported values, we adopted the same strategy and employed \textit{$\tau$} = \textit{$\sigma$}$_{exp}$/\textit{$\sigma$}$_{cal}$. The approximation provides the relaxation time of the order of 10$^{-16}$ s. Incorporating the calculated \textit{$\tau$} values at different temperatures into PF reported by R. He \textit{et al.} \cite{He2016}, a close agreement in the range of 300 - 700~K could be seen between calculated and reported PF/\textit{$\tau$} (Fig.~\ref{transport2}). However, at higher temperatures, the reported PF/\textit{$\tau$} deviates from our calculated values. But this is not a matter of concern since the trend of PF/\textit{$\tau$} is more important for us. Unfortunately, the PF/\textit{$\tau$} for Cr-group is almost half of the order of V-group. Nevertheless, we establish a nice agreement between our calculated and reported values for CoTiSb and CoNbSn system and noticed an appreciable enhancement in TE properties on doping. This motivates us to consider the importance of doping for V-group and Cr-group also. Thus, we proceed to the systems of our interest and explore the transport properties as a function of doping. 

\subsection{Effects of doping and temperature on transport properties} 
\begin{figure}
\centering
 \includegraphics[scale=0.40]{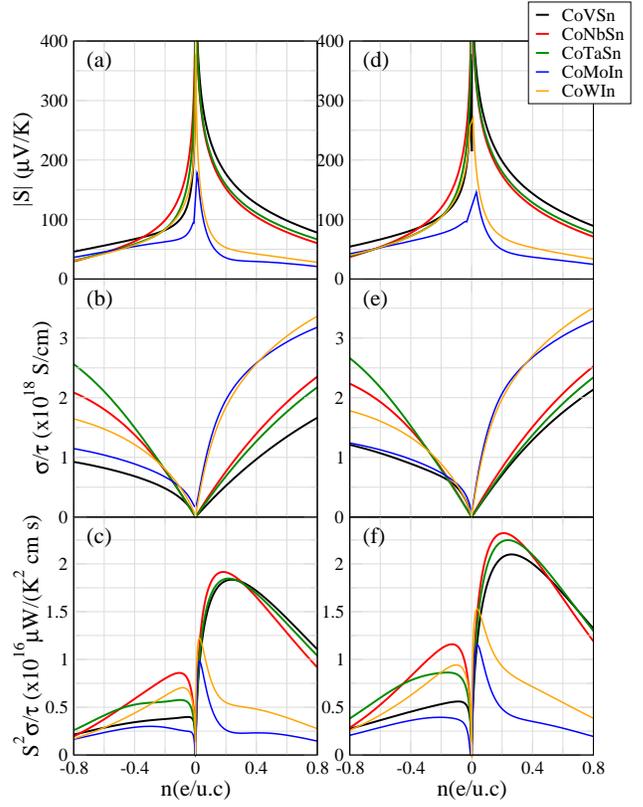}
 \caption{(a-c) shows the trend of Seebeck coefficient, electrical conductivity, and power factor as a function of doping at 700 K whereas (d-f) shows the trend of Seebeck coefficient, electrical conductivity, and power factor as a function of doping at 900 K. Power factor and electrical conductivity are plotted with respect to relaxation time.}
\label{doping}
\end{figure}

Fig.~\ref{doping} shows the transport properties of V-group and Cr-group at 700~K and 900~K, respectively. The trend of Seebeck coefficient and electrical conductivity is explained before. For all the systems at 700~K and 900~K, Seebeck coefficient tends to fall whereas electrical conductivity tends to improve at higher doping levels. The behavior of Seebeck coefficient is in quite good agreement with Mott equation and improving the carrier concentration will definitely enhance the electrical conductivity. Fig.~\ref{doping}(c) and Fig.~\ref{doping}(f) show the PF/\textit{$\tau$} as a function of doping at 700~K and 900~K, respectively. Again, the trend of PF/\textit{$\tau$} is same for all the systems. For all the temperatures, PF/\textit{$\tau$} increases substantially on both electron and hole doping and then drops at higher doping levels. However, hole doping dominates for all the systems. The maximum PF/\textit{$\tau$} is obtained near the band edge in all cases, owing to high Seebeck coefficient when the Fermi level is near the band edge. Table~\ref{tab2} lists the doping concentration (at which maximum PF/\textit{$\tau$} is obtained), maximum PF/\textit{$\tau$}, and corresponding Seebeck coefficient and electrical conductivity at 900~K. It is easy to see that the PF/\textit{$\tau$} for V-group is of the order of CoTiSb whereas the PF/\textit{$\tau$} of Cr-group is almost half as that of the CoTiSb. In fact, our calculations suggest that at varying doping levels, the PF/\textit{$\tau$} of V-group surpasses the PF/\textit{$\tau$} of CoTiSb. The maximum PF/\textit{$\tau$} is obtained for CoNbSn, closely followed by CoTaSn, and then CoVSn. Till now, experimentalists have focused on \textit{n}-type doping in CoNbSn and there is no available experimental evidence for \textit{p}-type doping. Recently in 2016, R. He \textit{et al.}, while exploring the TE properties for \textit{n}-type CoNbSn, broadly hinted that higher TE properties could be expected for \textit{p}-type doping\cite{He2016}. Here, we propose that CoNbSn could be a potential TE material on \textit{p}-type doping and is in agreement with previous theoretical prediction\cite{Yang2008}. We also predict that the recently discovered member of the hH family, CoTaSn, could be a promising \textit{p}-type TE material. The experimental realization of CoTaSn\cite{Zakutayev2013} and our theoretical predictions provide an interesting platform for experimentalists. CoVSn also shows promising TE properties on \textit{p}-type doping. Recently, David J. Singh and group also predicted the high TE performance of the \textit{p}-type doped CoVSn \cite{Shi2017}. C. S. Lue \textit{et al.} synthesized CoVSn with partial atomic disordering and suggested that CoVSn could be stable and can be produced at least with partial disordering\cite{Lue2001}. But unfortunately, there has been no study on CoVSn since then.

The Cr-group exhibits minimum PF/\textit{$\tau$} and this can be attributed to their band structure. The VBM of Cr-group is somewhat non-dispersed as compared to dispersed bands in VBM of V-group. Flat bands are an indication of weak interactions between atoms. The charge carriers in flat band region are acted upon by two or more atoms, thereby reducing their mobility. This leads to the heavier effective mass of charge carriers and high Seebeck coefficient.

\begin{table}[h]
\begin{tabular}{l|c|c|c|c}
\hline
Compound  & Doping    & S$^2\sigma$/$\tau$ ($\times$10$^{16}$-  & S  & $\sigma$/$\tau$ ($\times$10$^{18}$-  \\
	  & (e/u.c.)  & $\mu$W/K$^2$~cm~s) & ($\mu$~V/K)  & S/cm~s) \\
\hline
CoTiSb  & 0.22  & 2.02  & 179  & 0.63 \\ 
CoVSn   & 0.26  & 2.10  & 175  & 0.69 \\
CoNbSn  & 0.21  & 2.33  & 170  & 0.81 \\
CoTaSn  & 0.23  & 2.24  & 169  & 0.80 \\
CoMoIn  & 0.03  & 1.16  & 136  & 0.61 \\
CoWIn   & 0.03  & 1.52  & 170  & 0.54 \\
\hline
\end{tabular}
\caption{shows the optimum doping concentration at 900 K for CoTiSb, V-group, and Cr-group at which maximum power factor is obtained. Also at optimum doping concentration, power factor, Seebeck coefficient, and electrical conductivity are shown. Power factor and electrical conductivity are shown with respect to relaxation time.}
\label{tab2}
\end{table}

Note that our calculated PF values for different doping concentrations depend highly on relaxation time. The approximate value of relaxation time for CoTiSb, \textit{$\tau$} = 2 $\times$ 10$^{-16}$~s, which we used to compare the calculated PF/\textit{$\tau$} with reported PF/\textit{$\tau$}, cannot be utilized for doped CoTiSb. This is because the relaxation time may vary on doping the system. To validate our point, we have compared our calculated PF at 900~K for CoTiSb at 0.22 hole doping with T. Wu’s work\cite{Wu2007}. In this work, a maximum PF of 23 $\mu$W/K$^2$~cm at 850~K for CoTiSb at 0.15 hole doping was reported. Once again, employing the same strategy as before, we have approximated the relaxation time for doped CoTiSb of the order of 10$^{-15}$~s. Using this relaxation time, we obtain PF = 20.2 $\mu$W/K$^2$~cm at 900~K for 0.22 hole doping. This is close to the maximum PF ever achieved for doped CoTiSb system. We expect that the actual relaxation time for the V-group and Cr-group on hole doping would be of the same order \textit{i.e.} 10$^{-15}$~s. Assuming \textit{$\tau$} = 10$^{-15}$~s for doped V-group and Cr-group, we propose PF of 21, 23, 22, 11, and 15 $\mu$W/K$^2$~cm for CoVSn (at 0.26 \textit{p}-type doping), CoNbSn (at 0.21 \textit{p}-type doping), CoTaSn (at 0.23 \textit{p}-type doping), CoMoIn (at 0.03 \textit{p}-type doping), and CoWIn (at 0.03 \textit{p}-type doping), respectively.

\section{Discussion and Conclusions}
Till now, the main focus in 18 valence electron count (VEC) half-Heusler alloys is centered on \textit{M}NiSn and \textit{M}CoSb, where \textit{M} = Ti, Zr, Hf. The emphasis is on tailoring the properties of existing materials. However, the \textit{ZT} value is yet to see the progressive increment. This motivates finding new plausible 18-VEC hH alloys along with improving the TE properties of existing materials. The interplay between theory and experiment had been fruitful in designing new materials. In the present work, we utilized the similar theoretical approach for predicting new Co-based hH alloys and exploring the TE properties of existing ones. However, the theoretical prediction of new materials is relatively easier as compared to experimental realization. The manifesting challenges for experimentalists are to synthesize the ordered compositions and the vulnerability of hH alloys for anti-site disorder. The partial disordering and anti-site disorder may affect the properties significantly. 

Here, we have systematically investigated nine 18-VEC Co\textit{YZ} systems in \textit{F$\bar{4}$3m} and \textit{P6$_3$/mmc} symmetry. All 9 systems were found to be metallic in \textit{P6$_3$/mmc} symmetry and semiconductors, except CoCrIn, in \textit{F$\bar{4}$3m} symmetry. The V-group was found to have indirect band-gap whereas Cr-group possesses direct band-gap. Hence, CoMoIn and CoWIn are expected to possess low thermal conductivity than the other systems on the basis of direct band gap. The stability of V-group and Cr-group was confirmed by phonon calculations. The transport properties of V-group and Cr-group were found to be more promising on hole doping. We propose that the hole doped CoNbSn system could be a more promising thermoelectric material than the existing electron doped CoNbSn systems. Also, CoVSn, CoNbSn, and CoTaSn, at 900~K, shows higher power factor than the well-known CoTiSb on 0.26, 0.21, and 0.23 hole doping per unit cell, respectively. The values of thermopower for hole doped V-group and Cr-group ranges from 136 to 175 $\mu$V/K at 900~K. Through this work, we hope to motivate experimentalists to synthesize CoVSn, CoMoIn, CoWIn, and improve the TE properties of existing CoNbSn and CoTaSn by doping.

\section{Acknowledgement}
MZ is thankful to CSIR for granting senior research fellowship. Computations were performed on HP cluster at the Computer Center for Scientific Computing (ICC), IIT Roorkee and at IFW Dresden, Germany. We thank Ulrike Nitzsche and Navneet Gupta for technical assistance. HCK gratefully acknowledge finanical support from the FIG program of IIT Roorkee (Grant CMD/FIG/100596).


\end{document}